\documentclass[10pt]{article}
\usepackage{graphicx}
\usepackage{amsmath}
\usepackage{amssymb}
\usepackage{caption2}
\begin{document}
\begin{center}
{\large {\bf \sc{  The ground states and first radial excitations  of the vector tetraquark states with explicit P-waves  via  the QCD sum rules }}} \\[2mm]
Zhi-Gang  Wang \footnote{E-mail: zgwang@aliyun.com.  }     \\
 Department of Physics, North China Electric Power University, Baoding 071003, P. R. China
\end{center}

\begin{abstract}
In this work, we choose the diquark-antidiquark type four-quark currents with an explicit P-wave between the diquark and antidiquark pairs to study the ground states and first radial excitations of the hidden-charm tetraquark states with the quantum numbers $J^{PC}=1^{--}$. And we obtain  the lowest vector tetraquark masses and make possible assignments of the existing $Y$ states. There indeed exists a hidden-charm  tetraquark   state with the $J^{PC}=1^{--}$ at the energy about $4.75\,\rm{GeV}$ as the first radial excitation to account for the BESIII data.
 \end{abstract}

 PACS number: 12.39.Mk, 12.38.Lg

Key words: Tetraquark  state, QCD sum rules

\section{Introduction}
There have been observed a number of $Y$ states,
such as the $Y(4008)$ observed in the $J/\psi \pi^+\pi^-$ channel  \cite{Belle4008-0707},
the $Y(4220)$ and $Y(4320)$ observed in the $J/\psi \pi^+\pi^-$ channel  \cite{BESIII-4220-4320-1611},
the $Y(4230)$ observed in the $\omega\chi_{c0}$ channel  \cite{BES-2014-4230},
the $Y(4260)$ observed in the $J/\psi \pi^+\pi^-$ channel  \cite{BaBar4260-0506,Belle-0707,CLEO-0606},
the $Y(4320)$ observed in the $\psi^{\prime} \pi^+\pi^-$ channel  \cite{BaBar4360-0610057},
the $Y(4360)$ and $Y(4660)$ observed in the $\psi^{\prime} \pi^+\pi^-$ channel  \cite{Belle4660-0707-1,Belle4660-0707-2},
  the $Y(4390)$ observed in the  $ \pi^+\pi^- h_c$   channel  \cite{BES-Y4390},
 the $Y(4469)$ observed in the $ D^{*0}D^{*-}\pi^{+}$ channel  \cite{BESIII-Y4469-PRL},
 the $Y(4484)$ observed in the $ K^+K^-J/\psi$ channel  \cite{BESIII-Y4484-CPC},
 the $Y(4544)$ observed in the $\omega\chi_{c1}$ channel
\cite{BESIII-Y4544-2401},
  the $Y(4630)$ observed in the
 $ \Lambda_c^+ \Lambda_c^-$   channel \cite{Belle4630-0807},
the $Y(4710)$ observed in the $ K^+K^-J/\psi$  channel  \cite{Y4710-BESIII},
   the $Y(4790)$ observed in the $ D_{s}^{\ast+}D_{s}^{\ast-}$ channel \cite{Y4790-BESIII}.

Recently, the BESIII collaboration studied the processes  $e^+e^-\to\omega X(3872)$ and $\gamma X(3872)$  using data samples with an integrated luminosity of $4.5~\text{fb}^{-1}$ at center-of-mass energies from 4.66 to 4.95~GeV, and observed that the relatively large cross section for the $e^+e^-\to\omega X(3872)$ process is mainly attributed to the  enhancement around 4.75 GeV, which maybe  indicate a potential structure in the $e^+e^-\to\omega X(3872)$  cross section  \cite{Y4750-BESIII}. If the enhancement is confirmed in the future by enough experimental data, there maybe exist another $Y$ state, the $Y(4750)$.

Even if the $Y(4220)$, $Y(4230)$ and $Y(4260)$ are the same particle, the $Y(4320)$, $Y(4360)$ and $Y(4390)$ are the same particle, the   $Y(4469)$, $Y(4484)$, and $Y(4544)$ are the same particle $Y(4500)$, the $Y(4630)$ and $Y(4660)$ are the same particle, the $Y(4710)$ and $Y(4750)$ are the same particle, the $Y$ states are beyond the compatibility of the traditional  quark models, we have to introduce tetraquark states, molecular states, hybrid states, etc,  to make reasonable assignments \cite{HXChen-review-1601,RFLebed-review-1610,AEsposito-review-1611,
FKGuo-review-1705,AAli-review-1706,SLOlsen-review-1708,MNielsen-review-1812,
YRLiu-review-1903,CPShen-review-1907,MZLiu-review-2404}.

In Refs.\cite{WZG-NPB-cucd,WZG-NPB-cscs}, we take the scalar, pseudoscalar, axialvector, vector and tensor (anti)diquarks  as the elementary constituents   to construct  the  four-quark currents without introducing  explicit P-waves,  and investigate  the hidden-charm and hidden-charm-hidden-strange tetraquark states with the quantum numbers $J^{PC}=1^{--}$ and $1^{-+}$ in a comprehensive and consistent way via the QCD sum rules,  and revisit the assignments of the  $X/Y$ states in the hidden-charm tetraquark scenario, see Tables \ref{Assignments-Table-cqcq}-\ref{Assignments-Table-csss}, where the subscripts $S$, $P$, $V$ ($\widetilde{V}$) and $A$ ($\tilde{A}$) stand for the scalar, pseudoscalar,  vector and axialvector  (anti)diquarks, respectively.

\begin{table}
\begin{center}
\begin{tabular}{|c|c|c|c|c|c|c|c|c|}\hline\hline
  $Y_c$                                                                    & $J^{PC}$  & $M_Y (\rm{GeV})$  & Assignments          \\ \hline

$[uc]_{P}[\overline{dc}]_{A}-[uc]_{A}[\overline{dc}]_{P}$                  &$1^{--}$   &$4.66\pm0.07$      & ?\,\,$Y(4660)$          \\

$[uc]_{P}[\overline{dc}]_{A}+[uc]_{A}[\overline{dc}]_{P}$                  &$1^{-+}$   &$4.61\pm0.07$      &                     \\

$[uc]_{S}[\overline{dc}]_{V}+[uc]_{V}[\overline{dc}]_{S}$                  &$1^{--}$   &$4.35\pm0.08$      & ?\,\,$Y(4360/4390)$     \\

$[uc]_{S}[\overline{dc}]_{V}-[uc]_{V}[\overline{dc}]_{S}$                  &$1^{-+}$   &$4.66\pm0.09$      &                     \\

$[uc]_{\tilde{V}}[\overline{dc}]_{A}-[uc]_{A}[\overline{dc}]_{\tilde{V}}$  &$1^{--}$   &$4.53\pm0.07$      &                     ? $Y(4500)$\\

$[uc]_{\tilde{V}}[\overline{dc}]_{A}+[uc]_{A}[\overline{dc}]_{\tilde{V}}$  &$1^{-+}$   &$4.65\pm0.08$      &                     \\

$[uc]_{\tilde{A}}[\overline{dc}]_{V}+[uc]_{V}[\overline{dc}]_{\tilde{A}}$  &$1^{--}$   &$4.48\pm0.08$      &                     ? $Y(4500)$\\

$[uc]_{\tilde{A}}[\overline{dc}]_{V}-[uc]_{V}[\overline{dc}]_{\tilde{A}}$  &$1^{-+}$   &$4.55\pm0.07$      &                     \\

$[uc]_{S}[\overline{dc}]_{\tilde{V}}-[uc]_{\tilde{V}}[\overline{dc}]_{S}$  &$1^{--}$   &$4.50\pm0.09$      &                     ? $Y(4500)$\\

$[uc]_{S}[\overline{dc}]_{\tilde{V}}+[uc]_{\tilde{V}}[\overline{dc}]_{S}$  &$1^{-+}$   &$4.50\pm0.09$      &                     \\

$[uc]_{P}[\overline{dc}]_{\tilde{A}}-[uc]_{\tilde{A}}[\overline{dc}]_{P}$  &$1^{--}$   &$4.60\pm0.07$      &                     \\

$[uc]_{P}[\overline{dc}]_{\tilde{A}}+[uc]_{\tilde{A}}[\overline{dc}]_{P}$  &$1^{-+}$   &$4.61\pm0.08$      &                     \\

$[uc]_{A}[\overline{dc}]_{A}$                                              &$1^{--}$   &$4.69\pm0.08$      & ?\,\,$Y(4660)$      \\
\hline\hline
\end{tabular}
\end{center}
\caption{ The possible assignments of the  hidden-charm tetraquark states, the isospin limit is implied \cite{WZG-NPB-cucd}. }\label{Assignments-Table-cqcq}
\end{table}

\begin{table}
\begin{center}
\begin{tabular}{|c|c|c|c|c|c|c|c|c|}\hline\hline
 $Y_c$                                                                     &$J^{PC}$   &$M_Y (\rm{GeV})$   &Assignments            \\ \hline

$[sc]_{P}[\overline{sc}]_{A}-[sc]_{A}[\overline{sc}]_{P}$                  &$1^{--}$   &$4.80\pm0.08$    & ? $Y(4790)$ \\

$[sc]_{P}[\overline{sc}]_{A}+[sc]_{A}[\overline{sc}]_{P}$                  &$1^{-+}$   &$4.75\pm0.08$    &   \\

$[sc]_{S}[\overline{sc}]_{V}+[sc]_{V}[\overline{sc}]_{S}$                  &$1^{--}$   &$4.53\pm0.08$    & \\

$[sc]_{S}[\overline{sc}]_{V}-[sc]_{V}[\overline{sc}]_{S}$                  &$1^{-+}$   &$4.83\pm0.09$    &  \\

$[sc]_{\tilde{V}}[\overline{sc}]_{A}-[sc]_{A}[\overline{sc}]_{\tilde{V}}$  &$1^{--}$   &$4.70\pm0.08$    & ? $Y(4710)$ \\

$[sc]_{\tilde{V}}[\overline{sc}]_{A}+[sc]_{A}[\overline{sc}]_{\tilde{V}}$  &$1^{-+}$   &$4.81\pm0.09$    & \\

$[sc]_{\tilde{A}}[\overline{sc}]_{V}+[sc]_{V}[\overline{sc}]_{\tilde{A}}$  &$1^{--}$   &$4.65\pm0.08$    & ? $Y(4660)$     \\

$[sc]_{\tilde{A}}[\overline{sc}]_{V}-[sc]_{V}[\overline{sc}]_{\tilde{A}}$  &$1^{-+}$   &$4.71\pm0.08$    &   \\

$[sc]_{S}[\overline{sc}]_{\tilde{V}}-[sc]_{\tilde{V}}[\overline{sc}]_{S}$  &$1^{--}$   &$4.68\pm0.09$    & ? $Y(4660)$     \\

$[sc]_{S}[\overline{sc}]_{\tilde{V}}+[sc]_{\tilde{V}}[\overline{sc}]_{S}$  &$1^{-+}$   &$4.68\pm0.09$    &  ?? $X(4630)$ \\

$[sc]_{P}[\overline{sc}]_{\tilde{A}}-[sc]_{\tilde{A}}[\overline{sc}]_{P}$  &$1^{--}$   &$4.75\pm0.08$    &    \\

$[sc]_{P}[\overline{sc}]_{\tilde{A}}+[sc]_{\tilde{A}}[\overline{sc}]_{P}$  &$1^{-+}$   &$4.75\pm0.08$    &    \\

$[sc]_{A}[\overline{sc}]_{A}$                                              &$1^{--}$   &$4.85\pm0.09$    &     \\

\hline\hline
\end{tabular}
\end{center}
\caption{ The possible assignments of the hidden-charm-hidden-strange tetraquark states \cite{WZG-NPB-cscs}. }\label{Assignments-Table-csss}
\end{table}

In Refs.\cite{WZG-EPJC-P-4260,WZG-EPJC-P-wave}, we introduce an explicit P-wave between the diquark and antidiquark pairs  to construct the hidden-charm four-quark currents,
 and explore the  vector tetraquark states systematically via the QCD sum rules, and obtain the lowest vector tetraquark masses up to today, and revisit the assignments of the  $Y$ states in the  hidden-charm tetraquark scenario, see Table \ref{Assignments-Table-P-wave}, where the angular momenta $\vec{S}=\vec{S}_{qc}+\vec{S}_{\bar{q}\bar{c}}$ and $\vec{J}=\vec{L}+\vec{S}$.

\begin{table}
\begin{center}
\begin{tabular}{|c|c|c|c|c|c|c|c|}\hline\hline
$|S_{qc}, S_{\bar{q}\bar{c}}; S, L; J\rangle$                                &$M_Y(\rm{GeV})$     &Assignments    \\ \hline
$|0, 0; 0, 1; 1\rangle$                                                      &$4.24\pm0.10$       &$Y(4220)$           \\ \hline

$|1, 1; 0, 1; 1\rangle$                                                      &$4.28\pm0.10$       &$Y(4220/4320)$        \\ \hline

$\frac{1}{\sqrt{2}}\left(|1, 0; 1, 1; 1\rangle+|0, 1; 1, 1; 1\rangle\right)$ &$4.31\pm0.10$       &$Y(4320/4390)$        \\ \hline

$|1, 1; 2, 1; 1\rangle$                                                      &$4.33\pm0.10$       &$Y(4320/4390)$      \\ \hline \hline
\end{tabular}
\end{center}
\caption{ The possible assignments of the  hidden-charm tetraquark states with explicit P-waves, the isospin limit is implied \cite{WZG-EPJC-P-wave}. }\label{Assignments-Table-P-wave}
\end{table}

From Tables \ref{Assignments-Table-cqcq}-\ref{Assignments-Table-P-wave}, we can see explicitly that there are no rooms for the $Y(4008)$ and $Y(4750)$ in the vector hidden-charm  tetraquark scenario. If we take the $X(3872)$ and $Z_c(3900)$ as the lowest tetraquark states with the $J^{PC}=1^{++}$ and $1^{--}$, respectively \cite{Maiani-Z4430-1405,Narison-X3872,Nielsen-1401,WangHuangtao-2014-PRD,WangZG-Z4430-CTP}, the $Y(4008)$ cannot be assigned as a vector tetraquark state  due to the small mass splitting $\delta M\approx 100\,\rm{MeV}$. If the $Y(4220/4230/4260)$ can be assigned as the ground state vector tetraquark state, the lowest vector hidden-charm tetraquark state can be obtained by the QCD sum rules, the $Y(4750)$ can be assigned as its first radial excitation  according to the mass gap $M_{Y(4750)}-M_{Y(4260)}=0.51\,\rm{GeV}$, which happens to be our  naive expectation of the mass gap between the ground state and first radial excitation.

If the $Y(4750)$ can be assigned as the first radial excitation of the $Y(4220/4230/4260)$, there maybe exist a spectrum for the first radial excited states, which lie at about $4.8\,\rm{GeV}$, it is interesting to explore such a possibility.

As we know, the heavy-light diquarks $\varepsilon^{ijk} q^{T}_j C\Gamma Q_k$  have  five  structures, where the $i$, $j$ and $k$ are color indexes, $C\Gamma=C\gamma_5$, $C$, $C\gamma_\mu \gamma_5$,  $C\gamma_\mu $ and $C\sigma_{\mu\nu}$ for the scalar, pseudoscalar, vector, axialvector  and  tensor diquarks, respectively, the P-wave is implicitly embodied in the negative parity of the diquarks. We can also introduce an explicit P-wave inside  the heavy-light diquarks to obtain $\varepsilon^{ijk} q^{T}_j C\Gamma\stackrel{\leftrightarrow}{\partial}_\mu Q_k$, where  the derivative  $\stackrel{\leftrightarrow}{\partial}_\mu=\stackrel{\rightarrow}{\partial}_\mu-\stackrel{\leftarrow}{\partial}_\mu$ embodies  the explicit P-wave, then take the diquarks  $\varepsilon^{ijk} q^{T}_j C\Gamma\stackrel{\leftrightarrow}{\partial}_\mu Q_k$ as the basic building blocks to construct the four-quark currents to study the tetraquark states with the $J^{PC}=1^{--}$, and we would  explore such a possibility in our next work.

In Refs.\cite{WZG-NPB-cucd,WZG-NPB-cscs,WZG-EPJC-P-4260,WZG-EPJC-P-wave},
 we use the (modified) energy scale formula  to obtain the suitable  energy scales of the QCD spectral densities to enhance  the pole contributions and improve the convergent behaviors of the operator product expansion \cite{Wang-tetra-formula}, it is the unique feature of our works. In this direction,
we  have also explored  the hidden-charm tetraquark states with the $J^{PC}=0^{++}$,  $0^{-+}$, $0^{--}$,  $1^{+-}$, $2^{++}$ \cite{WZG-HC-spectrum-PRD,WZG-tetra-psedo-NPB}, hidden-bottom tetraquark states with the $J^{PC}=0^{++}$, $1^{+-}$, $2^{++}$ \cite{WZG-HB-spectrum-EPJC}, hidden-charm molecular states with the $J^{PC}=0^{++}$, $1^{+-}$, $2^{++}$ \cite{WZG-mole-IJMPA}, doubly-charm tetraquark (molecular) states with the $J^{P}=0^{+}$, $1^{+}$, $2^{+}$ \cite{WZG-tetra-cc-EPJC} (\cite{WZG-XQ-mole-EPJA}), hidden-charm pentaquark (molecular) states \cite{WZG-penta-cc-IJMPA-2050003}(\cite{XWWang-penta-mole}), and assign the existing exotic states consistently.

In the isospin limit, the vector tetraquark states with the symbol valence  quarks
\begin{eqnarray}
I=1&:& c\bar{c} u\bar{d}\, , \, \, \,  c\bar{c} \frac{u\bar{u}-d\bar{d}}{\sqrt{2}}\, , \, \, \, c\bar{c} d\bar{u}\, , \nonumber\\
I=0&:& c\bar{c} \frac{u\bar{u}+d\bar{d}}{\sqrt{2}}\, ,
\end{eqnarray}
have degenerated masses, and we would  explore the $c\bar{c} u\bar{d}$ tetraquark states for simplicity. We update the analysis of our previous works \cite{WZG-EPJC-P-4260,WZG-EPJC-P-wave}, extend our previous works to explore  the first radial citations of the vector hidden-charm  tetraquark states with the QCD sum rules systematically, and take the modified  energy scale formula to acquire the suitable   energy scales of the QCD spectral densities, and make  possible assignments of the existing $Y$ states and make predictions for the mass spectrum of the first radial excitations at the energy about $4.8\,\rm{GeV}$.

The article is arranged as follows:  we derive the QCD sum rules for  the vector   tetraquark states in section 2; in section 3, we   present the numerical results and discussions; section 4 is reserved for our conclusion.

\section{QCD sum rules for  the  vector tetraquark states}
We write down  the two-point correlation functions  $\Pi_{\mu\nu}(p)$ and $\Pi_{\mu\nu\alpha\beta}(p)$ firstly,
\begin{eqnarray}
\Pi_{\mu\nu}(p)&=&i\int d^4x e^{ip \cdot x} \langle0|T\left\{J_\mu(x)J_\nu^{\dagger}(0)\right\}|0\rangle \, , \\
\Pi_{\mu\nu\alpha\beta}(p)&=&i\int d^4x e^{ip \cdot x} \langle0|T\left\{J_{\mu\nu}(x)J_{\alpha\beta}^{\dagger}(0)\right\}|0\rangle \, ,
\end{eqnarray}
where $J_\mu(x)=J_\mu^1(x)$, $J_\mu^2(x)$ and $J_\mu^3(x)$,
\begin{eqnarray}
J^1_\mu(x)&=&\frac{\varepsilon^{ijk}\varepsilon^{imn}}{\sqrt{2}}u^{Tj}(x)C\gamma_5 c^k(x)\stackrel{\leftrightarrow}{\partial}_\mu \bar{d}^m(x)\gamma_5 C \bar{c}^{Tn}(x) \, ,
\end{eqnarray}
\begin{eqnarray}
J^2_\mu(x)&=&\frac{\varepsilon^{ijk}\varepsilon^{imn}}{\sqrt{2}}u^{Tj}(x)C\gamma_\alpha c^k(x)\stackrel{\leftrightarrow}{\partial}_\mu \bar{d}^m(x)\gamma^\alpha C \bar{c}^{Tn}(x) \, ,
\end{eqnarray}
\begin{eqnarray}
J^3_\mu(x)&=&\frac{\varepsilon^{ijk}\varepsilon^{imn}}{2}\left[u^{Tj}(x)C\gamma_\mu c^k(x)\stackrel{\leftrightarrow}{\partial}_\alpha \bar{d}^m(x)\gamma^\alpha C \bar{c}^{Tn}(x) \right.\nonumber\\
&&\left.+u^{Tj}(x)C\gamma^\alpha c^k(x)\stackrel{\leftrightarrow}{\partial}_\alpha \bar{d}^m(x)\gamma_\mu C \bar{c}^{Tn}(x)\right]\, ,
\end{eqnarray}
\begin{eqnarray}
J_{\mu\nu}(x)&=&\frac{\varepsilon^{ijk}\varepsilon^{imn}}{2\sqrt{2}}\left[u^{Tj}(x)C\gamma_5 c^k(x)\stackrel{\leftrightarrow}{\partial}_\mu \bar{d}^m(x)\gamma_\nu C \bar{c}^{Tn}(x) \right.\nonumber\\
&&+u^{Tj}(x)C\gamma_\nu c^k(x)\stackrel{\leftrightarrow}{\partial}_\mu \bar{d}^m(x)\gamma_5 C \bar{c}^{Tn}(x) \nonumber  \\
&&-u^{Tj}(x)C\gamma_5 c^k(x)\stackrel{\leftrightarrow}{\partial}_\nu \bar{d}^m(x)\gamma_\mu C \bar{c}^{Tn}(x) \nonumber \\
&&\left.-u^{Tj}(x)C\gamma_\mu c^k(x)\stackrel{\leftrightarrow}{\partial}_\nu \bar{d}^m(x)\gamma_5 C \bar{c}^{Tn}(x)\right]\, .
\end{eqnarray}
Under charge conjugation transform $\widehat{C}$, the currents $J_\mu(x)$ and $J_{\mu\nu}(x)$ have the properties,
\begin{eqnarray}
\widehat{C}J_{\mu}(x)\widehat{C}^{-1}&=&- J_{\mu}(x) \, , \nonumber\\
\widehat{C}J_{\mu\nu}(x)\widehat{C}^{-1}&=&- J_{\mu\nu}(x) \, ,
\end{eqnarray}
in other words, they have negative conjugation.

At the hadron side, we insert  a complete set of intermediate hadronic states with
the same quantum numbers as the interpolating currents $J_\mu(x)$ and $J_{\mu\nu}(x)$ into the
correlation functions  $\Pi_{\mu\nu}(p)$ and $\Pi_{\mu\nu\alpha\beta}(p)$ respectively   to acquire  the hadronic representation
\cite{SVZ79-1,SVZ79-2,Reinders85}. Then we isolate the ground states and obtain the expressions,
\begin{eqnarray}
\Pi_{\mu\nu}(p)&=&\frac{\lambda_{Y}^2}{M_{Y}^2-p^2}\left(-g_{\mu\nu} +\frac{p_\mu p_\nu}{p^2}\right) + \cdots \, \, ,\nonumber\\
&=&\Pi_Y(p^2)\left(-g_{\mu\nu} +\frac{p_\mu p_\nu}{p^2}\right) +\cdots  \, , \\
\Pi_{\mu\nu\alpha\beta}(p)&=&\frac{\lambda_{ Y}^2}{M_{Y}^2\left(M_{Y}^2-p^2\right)}\left(p^2g_{\mu\alpha}g_{\nu\beta} -p^2g_{\mu\beta}g_{\nu\alpha} -g_{\mu\alpha}p_{\nu}p_{\beta}-g_{\nu\beta}p_{\mu}p_{\alpha}+g_{\mu\beta}p_{\nu}p_{\alpha}+g_{\nu\alpha}p_{\mu}p_{\beta}\right) \nonumber\\
&&+\frac{\lambda_{ Z}^2}{M_{Z}^2\left(M_{Z}^2-p^2\right)}\left( -g_{\mu\alpha}p_{\nu}p_{\beta}-g_{\nu\beta}p_{\mu}p_{\alpha}+g_{\mu\beta}p_{\nu}p_{\alpha}+g_{\nu\alpha}p_{\mu}p_{\beta}\right) +\cdots \, , \nonumber\\
&=&\widetilde{\Pi}_{Y}(p^2)\left(p^2g_{\mu\alpha}g_{\nu\beta} -p^2g_{\mu\beta}g_{\nu\alpha} -g_{\mu\alpha}p_{\nu}p_{\beta}-g_{\nu\beta}p_{\mu}p_{\alpha}+g_{\mu\beta}p_{\nu}p_{\alpha}+g_{\nu\alpha}p_{\mu}p_{\beta}\right) \nonumber\\
&&+\widetilde{\Pi}_{Z}(p^2)\left( -g_{\mu\alpha}p_{\nu}p_{\beta}-g_{\nu\beta}p_{\mu}p_{\alpha}+g_{\mu\beta}p_{\nu}p_{\alpha}+g_{\nu\alpha}p_{\mu}p_{\beta}\right) \, ,
\end{eqnarray}
where we take the definitions for the pole residues  $\lambda_{Y}$ and $\lambda_{Z}$,
\begin{eqnarray}
\langle 0|J_\mu(0)|Y(p)\rangle &=&\lambda_{Y} \,\varepsilon_\mu \, , \nonumber\\
  \langle 0|J_{\mu\nu}(0)|Y(p)\rangle &=& \frac{\lambda_{Y}}{M_{Y}} \, \varepsilon_{\mu\nu\alpha\beta} \, \varepsilon^{\alpha}p^{\beta}\, , \nonumber\\
 \langle 0|J_{\mu\nu}(0)|Z(p)\rangle &=& \frac{\lambda_{Z}}{M_{Z}} \left(\varepsilon_{\mu}p_{\nu}-\varepsilon_{\nu}p_{\mu} \right)\, ,
\end{eqnarray}
the $\varepsilon_\mu$ are the polarization vectors  of the   tetraquark states $Y$ and $Z$ with the quantum numbers $J^{PC}=1^{--}$ and $1^{+-}$, respectively.
Now we project out the components $\Pi_{Y}(p^2)$ and $\Pi_{Z}(p^2)$ explicitly with the projectors  $P_{Y}^{\mu\nu\alpha\beta}$ and $P_{Z}^{\mu\nu\alpha\beta}$,
\begin{eqnarray}
\Pi_{Y}(p^2)&=&p^2\widetilde{\Pi}_{Y}(p^2)=P_{Y}^{\mu\nu\alpha\beta}\Pi_{\mu\nu\alpha\beta}(p) \, , \nonumber\\
\Pi_{Z}(p^2)&=&p^2\widetilde{\Pi}_{Z}(p^2)=P_{Z}^{\mu\nu\alpha\beta}\Pi_{\mu\nu\alpha\beta}(p) \, ,
\end{eqnarray}
where
\begin{eqnarray}
P_{Y}^{\mu\nu\alpha\beta}&=&\frac{1}{6}\left( g^{\mu\alpha}-\frac{p^\mu p^\alpha}{p^2}\right)\left( g^{\nu\beta}-\frac{p^\nu p^\beta}{p^2}\right)\, , \nonumber\\
P_{Z}^{\mu\nu\alpha\beta}&=&\frac{1}{6}\left( g^{\mu\alpha}-\frac{p^\mu p^\alpha}{p^2}\right)\left( g^{\nu\beta}-\frac{p^\nu p^\beta}{p^2}\right)-\frac{1}{6}g^{\mu\alpha}g^{\nu\beta}\, ,
\end{eqnarray}
 we take the components $\Pi_{Y}(p^2)$ as we explore the hidden-charm tetraquark states with the $J^{PC}=1^{--}$.

 At the QCD side, we accomplish  the operator product expansion up to the vacuum condensates of   dimension-10, and consider the vacuum condensates which are vacuum expectations  of the operators  of the orders $\mathcal{O}( \alpha_s^{k})$ with $k\leq 1$ consistently, i.e. we consider the
 $ \langle\bar{q}q\rangle$, $ \langle\frac{\alpha_{s}GG}{\pi}\rangle$,
 $\langle\bar{q}g_s\sigma Gq\rangle$,
$ \langle\bar{q}q\rangle^2$,
$\langle\bar{q}q\rangle\langle\frac{\alpha_{s}GG}{\pi}\rangle$,
$\langle\bar{q}q\rangle\langle\bar{q}g_s\sigma Gq\rangle$,
$\langle\bar{q}g_s\sigma Gq\rangle^2$ and  $\langle\bar{q}q\rangle^2\langle\frac{\alpha_{s}GG}{\pi}\rangle$. The interested readers can consult Refs.\cite{WZG-EPJC-P-4260,WZG-EPJC-P-wave} for more details.

Now  we adopt quark-hadron duality below the continuum thresholds  $s_0$  and $s_0^\prime$, respectively,  and accomplish  Borel transform  with respect  to
 $P^2=-p^2$ to acquire  two QCD sum rules:
\begin{eqnarray}\label{QCDST-1S}
\lambda^2_{Y}\, \exp\left(-\frac{M^2_{Y}}{T^2}\right)&=& \int_{4m_c^2}^{s_0} ds  \,\rho_{QCD}(s)  \exp\left(-\frac{s}{T^2}\right) \, ,
\end{eqnarray}
\begin{eqnarray}\label{QCDST-2S}
\lambda^2_{Y}\, \exp\left(-\frac{M^2_{Y}}{T^2}\right)+\lambda^2_{Y^\prime}\, \exp\left(-\frac{M^2_{Y^\prime}}{T^2}\right)&=& \int_{4m_c^2}^{s^\prime_0} ds  \,\rho_{QCD}(s) \exp\left(-\frac{s}{T^2}\right) \, ,
\end{eqnarray}
where the  $\rho_{QCD}(s) $ are the QCD spectral densities obtained through dispersion relation, the $s_0$  and $s_0^\prime$ correspond to the ground states $Y$ and first radial excitations $Y^\prime$, respectively.

 We  adopt   the notations $\tau=\frac{1}{T^2}$, $D^n=\left( -\frac{d}{d\tau}\right)^n$, and  use  the subscripts $1$ and $2$ to denote  the  $Y$ and  $Y^\prime$ respectively to simplify the expressions.
 Now we rewrite the two QCD sum rules in Eqs.\eqref{QCDST-1S}-\eqref{QCDST-2S} as
\begin{eqnarray}\label{QCDSR-I}
\lambda_1^2\exp\left(-\tau M_1^2 \right)&=&\Pi_{QCD}(\tau) \, ,
\end{eqnarray}
\begin{eqnarray}\label{QCDSR-II-re}
\lambda_1^2\exp\left(-\tau M_1^2 \right)+\lambda_2^2\exp\left(-\tau M_2^2 \right)&=&\Pi^{\prime}_{QCD}(\tau) \, ,
\end{eqnarray}
where the $\Pi_{QCD}(\tau)$ and $\Pi^{\prime}_{QCD}(\tau)$  represent the correlation functions   below the continuum thresholds $s_0$ and $s_0^\prime$, respectively. We derive the QCD sum rules in Eq.\eqref{QCDSR-I} with respect   to $\tau$ to get the  ground states masses,
\begin{eqnarray}\label{QCDSR-I-Dr}
M_1^2&=&\frac{D\Pi_{QCD}(\tau)}{\Pi_{QCD}(\tau)}\, ,
\end{eqnarray}
then it is straightforward to get  the ground state masses and pole residues with the two coupled QCD sum rules, see Eq.\eqref{QCDSR-I} and Eq.\eqref{QCDSR-I-Dr} \cite{WZG-EPJC-P-4260,WZG-EPJC-P-wave}.

Then we  derive  the QCD sum rules in Eq.\eqref{QCDSR-II-re} with respect  to $\tau$ to get
\begin{eqnarray}\label{QCDSR-II-Dr}
\lambda_1^2M_1^2\exp\left(-\tau M_1^2 \right)+\lambda_2^2M_2^2\exp\left(-\tau M_2^2 \right)&=&D\Pi^{\prime}_{QCD}(\tau) \, .
\end{eqnarray}
From Eq.\eqref{QCDSR-II-re} and Eq.\eqref{QCDSR-II-Dr}, we get   the QCD sum rules,
\begin{eqnarray}\label{QCDSR-II-Residue}
\lambda_i^2\exp\left(-\tau M_i^2 \right)&=&\frac{\left(D-M_j^2\right)\Pi^{\prime}_{QCD}(\tau)}{M_i^2-M_j^2} \, ,
\end{eqnarray}
where  $i \neq j$.
Then we derive  the QCD sum rules in Eq.\eqref{QCDSR-II-Residue} with respect  to $\tau$ to acquire
\begin{eqnarray}
M_i^2&=&\frac{\left(D^2-M_j^2D\right)\Pi_{QCD}^{\prime}(\tau)}{\left(D-M_j^2\right)\Pi_{QCD}^{\prime}(\tau)} \, , \nonumber\\
M_i^4&=&\frac{\left(D^3-M_j^2D^2\right)\Pi_{QCD}^{\prime}(\tau)}{\left(D-M_j^2\right)\Pi_{QCD}^{\prime}(\tau)}\, .
\end{eqnarray}
 The squared masses $M_i^2$ obey  the  equation,
\begin{eqnarray}
M_i^4-b M_i^2+c&=&0\, ,
\end{eqnarray}
where
\begin{eqnarray}
b&=&\frac{D^3\otimes D^0-D^2\otimes D}{D^2\otimes D^0-D\otimes D}\, , \nonumber\\
c&=&\frac{D^3\otimes D-D^2\otimes D^2}{D^2\otimes D^0-D\otimes D}\, , \nonumber\\
D^j \otimes D^k&=&D^j\Pi^{\prime}_{QCD}(\tau) \,  D^k\Pi^{\prime}_{QCD}(\tau)\, ,
\end{eqnarray}
the subscripts $i=1,2$ and the superscripts  $j,k=0,1,2,3$.
At last, we solve the simple equation and obtain two solutions, i.e. the masses of the ground states and first radial excitations \cite{WangZG-Z4430-CTP,Baxi-G,WangZG-4430},
\begin{eqnarray}\label{QCDSR-II-M1}
M_1^2&=&\frac{b-\sqrt{b^2-4c} }{2} \, ,
\end{eqnarray}
\begin{eqnarray}\label{QCDSR-II-M2}
M_2^2&=&\frac{b+\sqrt{b^2-4c} }{2} \, .
\end{eqnarray}
We can acquire the ground state masses either from the QCD sum rules in Eq.\eqref{QCDSR-I-Dr} or in Eq.\eqref{QCDSR-II-M1}, but we prefer the QCD sum rules in Eq.\eqref{QCDSR-I-Dr}, because there are larger ground state contributions and less uncertainties from the continuum threshold parameters. We acquire  the masses and pole residues of the first radial excitations from the two coupled QCD sum rules in Eq.\eqref{QCDSR-II-Residue} and in Eq.\eqref{QCDSR-II-M2}.

\section{Numerical results and discussions}
We adopt  the traditional vacuum condensates $\langle
\bar{q}q \rangle=-(0.24\pm 0.01\, \rm{GeV})^3$,   $\langle
\bar{q}g_s\sigma G q \rangle=m_0^2\langle \bar{q}q \rangle$,
$m_0^2=(0.8 \pm 0.1)\,\rm{GeV}^2$,  $\langle \frac{\alpha_s
GG}{\pi}\rangle=0.012\pm0.003\,\rm{GeV}^4 $  at the energy scale  $\mu=1\, \rm{GeV}$
\cite{SVZ79-1,SVZ79-2,Reinders85,Colangelo-Review}, and choose the modified minimum subtracted mass $m_{c}(m_c)=(1.275\pm0.025)\,\rm{GeV}$ from the Particle Data Group \cite{PDG}, and set $m_u=m_d=0$. Moreover, we consider  the energy scale dependence of the input parameters,
\begin{eqnarray}
\langle\bar{q}q \rangle(\mu)&=&\langle\bar{q}q \rangle({\rm 1GeV})\left[\frac{\alpha_{s}({\rm 1GeV})}{\alpha_{s}(\mu)}\right]^{\frac{12}{33-2n_f}}\, , \nonumber\\
 \langle\bar{q}g_s \sigma Gq \rangle(\mu)&=&\langle\bar{q}g_s \sigma G q \rangle({\rm 1GeV})\left[\frac{\alpha_{s}({\rm 1GeV})}{\alpha_{s}(\mu)}\right]^{\frac{2}{33-2n_f}}\, , \nonumber\\
 m_c(\mu)&=&m_c(m_c)\left[\frac{\alpha_{s}(\mu)}{\alpha_{s}(m_c)}\right]^{\frac{12}{33-2n_f}} \, ,\nonumber\\
\alpha_s(\mu)&=&\frac{1}{b_0t}\left[1-\frac{b_1}{b_0^2}\frac{\log t}{t} +\frac{b_1^2(\log^2{t}-\log{t}-1)+b_0b_2}{b_0^4t^2}\right]\, ,
\end{eqnarray}
 where $t=\log \frac{\mu^2}{\Lambda_{QCD}^2}$, $b_0=\frac{33-2n_f}{12\pi}$, $b_1=\frac{153-19n_f}{24\pi^2}$, $b_2=\frac{2857-\frac{5033}{9}n_f+\frac{325}{27}n_f^2}{128\pi^3}$,  $\Lambda_{QCD}=210\,\rm{MeV}$, $292\,\rm{MeV}$  and  $332\,\rm{MeV}$ for the flavors  $n_f=5$, $4$ and $3$, respectively  \cite{PDG,Narison-mix}. We choose the flavor $n_f=4$ because we explore the tetraquark  states consist of the valence quarks $u$, $d$ and $c$. We evolve all the input parameters to the suitable  energy scales  $\mu$ to extract the masses  of the
   hidden-charm tetraquark states, which satisfy the modified energy scale formula  $\mu=\sqrt{M^2_{X/Y/Z}-(2{\mathbb{M}}_c+0.5\,\rm{GeV})^2}
   =\sqrt{M^2_{X/Y/Z}-(4.1\,\rm{GeV})^2}$, where the $\mathbb{M}$ is the effective charm quark mass \cite{WZG-EPJC-P-4260,WZG-EPJC-P-wave}.

In the scenario of tetraquark states, we can tentatively assign the $X(3915)$ and $X(4500)$ to be the 1S and 2S  states with the $J^{PC}=0^{++}$ \cite{X4140-tetraquark-Lebed,X3915-X4500-EPJC-WZG}, assign
the $Z_c(3900)$ and $Z_c(4430)$   to be   the 1S and 2S  states with the $J^{PC}=1^{+-}$, respectively \cite{Maiani-Z4430-1405,Nielsen-1401,WangZG-Z4430-CTP},   assign the $Z_c(4020)$ and $Z_c(4600)$ to be the 1S and 2S states with the $J^{PC}=1^{+-}$, respectively  \cite{WangZG-4430,ChenHX-Z4600-A}, and assign the $X(4140)$ and $X(4685)$ to be the 1S and 2S  states with the $J^{PC}=1^{++}$, respectively \cite{WZG-Di-X4140-EPJC,WZG-X4140-X4685}, where the energy  gaps between the 1S and 2S states  are about $0.57\sim 0.59 \,\rm{GeV}$.

In Refs.\cite{WZG-EPJC-P-4260,WZG-EPJC-P-wave}, we  choose the continuum threshold parameters as $\sqrt{s_0}=M_Y+0.55\sim0.60 \pm 0.10\,\rm{GeV}$ for the hidden-charm tetraquark states with the $J^{PC}=1^{--}$, the ground state contribution can be as large as $(49-81)\%$. Compared with the usually chosen pole contributions $(40-60)\%$ \cite{WZG-HC-spectrum-PRD,WZG-tetra-psedo-NPB,WZG-HB-spectrum-EPJC,WZG-mole-IJMPA,
WZG-tetra-cc-EPJC,WZG-XQ-mole-EPJA,WZG-penta-cc-IJMPA-2050003,XWWang-penta-mole}, the ground state contributions $(49-81)\%$ in the old analysis  are  too large in the QCD sum rules for the multiquark states, which maybe suffer from  contaminations from the first radial excitations, so in the  present calculations,
 we choose  slightly smaller continuum threshold  parameters $\sqrt{s_0}=M_Y+0.50\sim0.55\pm 0.10\,\rm{GeV}$ to reduce the ground state contributions, and perform a consistent and detailed analysis. Furthermore, the continuum threshold parameters $s_0^\prime$ are set to be    $\sqrt{s^\prime_0}=M_{Y^\prime}+0.40\pm 0.10\,\rm{GeV}$ according to the mass-gap of the $\psi^\prime$ and $\psi^{\prime \prime}$ from the Particle Data Group \cite{PDG}.

We search for the suitable  Borel parameters and continuum threshold parameters via trial and error.
The pole contributions (PC) and the vacuum condensate contributions ($D(n)$) are defined by
\begin{eqnarray}
{\rm{PC}}&=&\frac{\int_{4m_{c}^{2}}^{s_{0}/s_0^\prime}ds\rho_{QCD}\left(s\right)\exp\left(-\frac{s}{T^{2}}\right)} {\int_{4m_{c}^{2}}^{\infty}ds\rho_{QCD}\left(s\right)\exp\left(-\frac{s}{T^{2}}\right)}\, ,
\end{eqnarray}
 and
\begin{eqnarray}
D(n)&=&\frac{\int_{4m_{c}^{2}}^{s_{0}/s_0^\prime}ds\rho_{QCD,n}(s)\exp\left(-\frac{s}{T^{2}}\right)}
{\int_{4m_{c}^{2}}^{s_{0}/s_0^\prime}ds\rho_{QCD}\left(s\right)\exp\left(-\frac{s}{T^{2}}\right)}\, ,
\end{eqnarray}
respectively.

At last,  we obtain the Borel windows, continuum threshold parameters, suitable energy scales and  pole contributions, which are shown explicitly in Tables \ref{Borel-1S}-\ref{Borel-1S-2S}.
From the tables,  we can see explicitly that the pole contributions of the ground states (the ground states plus first radial excited states) are about $(40-60)\%$ ($(67-85)\%$), just like in our previous works on other tetraquark states \cite{WZG-HC-spectrum-PRD,WZG-tetra-psedo-NPB,WZG-HB-spectrum-EPJC,WZG-mole-IJMPA,
WZG-tetra-cc-EPJC,WZG-XQ-mole-EPJA,WZG-penta-cc-IJMPA-2050003,XWWang-penta-mole},    the pole dominance
criterion is satisfied very good. On the other hand, the contributions from the highest dimensional condensates play a minor important role,   $|D(10)|< 3\%$ or $\ll 1\%$ ($< 1\%$ or $\ll 1\%$) for the ground states (the ground states plus first radial excited states), the operator product  expansion converges  very good and better than that in our previous work \cite{WZG-EPJC-P-wave}. So we have confidence to extract reliable tetraquark masses and pole residues.

From Tables \ref{Borel-1S}-\ref{mass-2S}, we can see explicitly that the modified energy scale formula  can be well satisfied, and the relations  $\sqrt{s_0}=M_Y+0.50\sim0.55\pm 0.10\,\rm{GeV}$ and $\sqrt{s^\prime_0}=M_{Y^\prime}+0.40\pm 0.10\,\rm{GeV}$ are hold, our analysis is consistent.

\begin{figure}
 \centering
 \includegraphics[totalheight=5cm,width=7cm]{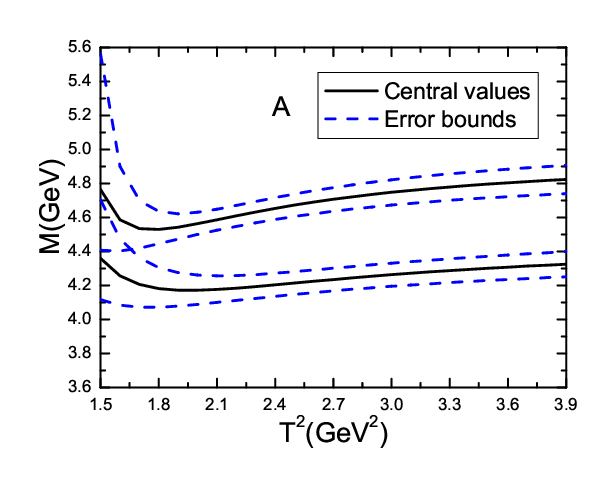}
 \includegraphics[totalheight=5cm,width=7cm]{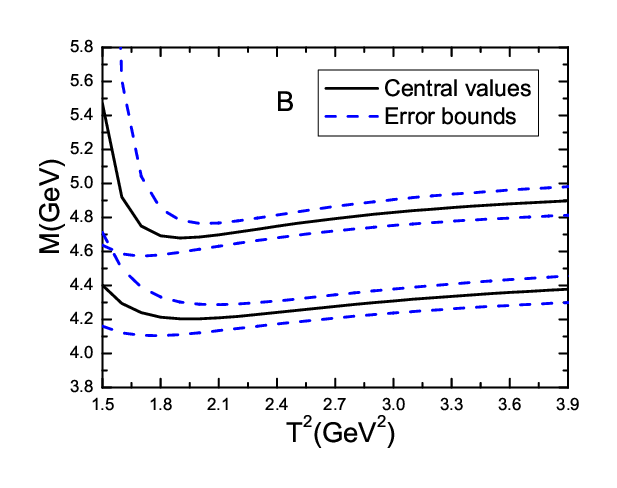}
 \includegraphics[totalheight=5cm,width=7cm]{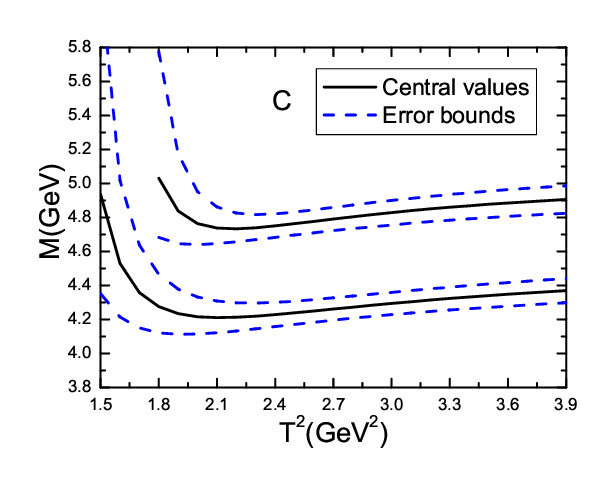}
 \includegraphics[totalheight=5cm,width=7cm]{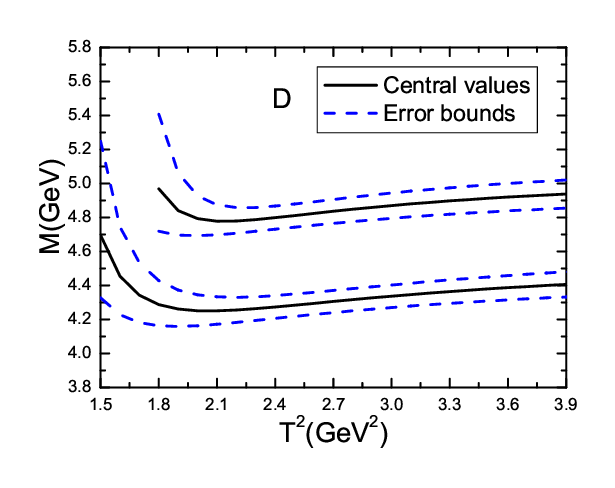}
  \caption{ The masses of the vector tetraquark states with variations of the Borel parameters $T^2$, where  the $A$, $B$, $C$ and $D$ stand for  the
  $|0, 0; 0, 1; 1\rangle$, $|1, 1; 0, 1; 1\rangle$, $\frac{1}{\sqrt{2}}\left(|1, 0; 1, 1; 1\rangle+|0, 1; 1, 1; 1\rangle\right)$ and $|1, 1; 2, 1; 1\rangle$  states,  respectively; the lower lines and upper lines stand for  the ground states and first radial excitations, respectively.   }\label{fig-mass-1S-2S}
\end{figure}

In Fig.\ref{fig-mass-1S-2S}, we plot the masses  of the ground states and first radial excitations of the  hidden-charm tetraquark states with the quantum numbers $J^{PC}=1^{--}$.
From the figure, we can see explicitly  that  there appear flat platforms in the Borel windows, the uncertainties come from the Borel parameters are rather small.

\begin{table}
\begin{center}
\begin{tabular}{|c|c|c|c|c|c|c|c|}\hline\hline
$|S_{qc}, S_{\bar{q}\bar{c}}; S, L; J\rangle$                                &$\mu(\rm{GeV})$ &$T^2(\rm{GeV}^2)$ &$\sqrt{s_0}(\rm{GeV})$ &pole    &$D(10)$ \\ \hline

$|0, 0; 0, 1; 1\rangle$                                                      &$1.1$         &$2.6-3.0$       &$4.75\pm0.10$        &$(40-65)\%$   &$< 1\%$ \\ \hline

$|1, 1; 0, 1; 1\rangle$                                                      &$1.2$         &$2.5-2.9$       &$4.80\pm0.10$        &$(39-64)\%$   &$< 3\%$ \\ \hline

$\frac{1}{\sqrt{2}}\left(|1, 0; 1, 1; 1\rangle+|0, 1; 1, 1; 1\rangle\right)$ &$1.3$         &$3.0-3.4$       &$4.85\pm0.10$        &$(38-60)\%$   &$\ll 1\%$ \\ \hline

$|1, 1; 2, 1; 1\rangle$                                                      &$1.3$         &$2.7-3.1$       &$4.85\pm0.10$        &$(39-63)\%$   &$< 1\%$ \\ \hline \hline
\end{tabular}
\end{center}
\caption{ The Borel windows $T^2$, continuum threshold parameters $s_0$, energy scales of the QCD spectral densities, contributions of the ground states, and values of the  $D(10)$.   }\label{Borel-1S}
\end{table}

\begin{table}
\begin{center}
\begin{tabular}{|c|c|c|c|c|c|c|c|}\hline\hline
$|S_{qc}, S_{\bar{q}\bar{c}}; S, L; J\rangle$                                &$\mu(\rm{GeV})$ &$T^2(\rm{GeV}^2)$ &$\sqrt{s^\prime_0}(\rm{GeV})$ &pole    &$D(10)$ \\ \hline

$|0, 0; 0, 1; 1\rangle$                                                      &$2.4$         &$2.8-3.2$       &$5.15\pm0.10$        &$(67-85)\%$   &$\ll 1\%$ \\ \hline

$|1, 1; 0, 1; 1\rangle$                                                      &$2.5$         &$2.6-3.0$       &$5.20\pm0.10$        &$(67-86)\%$   &$< 1\%$ \\ \hline

$\frac{1}{\sqrt{2}}\left(|1, 0; 1, 1; 1\rangle+|0, 1; 1, 1; 1\rangle\right)$ &$2.6$         &$3.0-3.4$       &$5.25\pm0.10$        &$(67-84)\%$   &$\ll 1\%$ \\ \hline

$|1, 1; 2, 1; 1\rangle$                                                      &$2.6$         &$2.7-3.1$       &$5.25\pm0.10$        &$(68-87)\%$   &$\ll 1\%$ \\ \hline \hline
\end{tabular}
\end{center}
\caption{ The Borel windows $T^2$, continuum threshold parameters $s_0^\prime$, energy scales of the QCD spectral densities, contributions of the ground states plus first radial excitations, and values of the  $D(10)$.   }\label{Borel-1S-2S}
\end{table}

\begin{table}
\begin{center}
\begin{tabular}{|c|c|c|c|c|c|c|c|}\hline\hline
$|S_{qc}, S_{\bar{q}\bar{c}}; S, L; J\rangle$                                &$M_Y(\rm{GeV})$  &$\lambda_Y(10^{-2}\rm{GeV}^6)$   \\ \hline

$|0, 0; 0, 1; 1\rangle$                                                      &$4.24\pm0.09$    &$2.28 \pm0.42$            \\ \hline

$|1, 1; 0, 1; 1\rangle$                                                      &$4.28\pm0.09$    &$4.80 \pm0.95$              \\ \hline

$\frac{1}{\sqrt{2}}\left(|1, 0; 1, 1; 1\rangle+|0, 1; 1, 1; 1\rangle\right)$ &$4.31\pm0.09$    &$2.94 \pm0.50$               \\ \hline

$|1, 1; 2, 1; 1\rangle$                                                      &$4.33\pm0.09$    &$6.55 \pm1.19$               \\ \hline \hline
\end{tabular}
\end{center}
\caption{ The masses and pole residues of the ground states.   }\label{mass-1S}
\end{table}

\begin{table}
\begin{center}
\begin{tabular}{|c|c|c|c|c|c|c|c|}\hline\hline
$|S_{qc}, S_{\bar{q}\bar{c}}; S, L; J\rangle$                                &$M_Y(\rm{GeV})$  &$\lambda_Y(10^{-2}\rm{GeV}^6)$   \\ \hline

$|0, 0; 0, 1; 1\rangle$                                                      &$4.75\pm0.10$    &$8.19 \pm1.23$            \\ \hline

$|1, 1; 0, 1; 1\rangle$                                                      &$4.81\pm0.10$    &$18.3 \pm3.0$              \\ \hline

$\frac{1}{\sqrt{2}}\left(|1, 0; 1, 1; 1\rangle+|0, 1; 1, 1; 1\rangle\right)$ &$4.85\pm0.09$    &$8.63 \pm1.22$               \\ \hline

$|1, 1; 2, 1; 1\rangle$                                                      &$4.86\pm0.10$    &$21.7 \pm3.4$               \\ \hline \hline
\end{tabular}
\end{center}
\caption{ The masses and pole residues of the first radial excited  states.   }\label{mass-2S}
\end{table}

\begin{table}
\begin{center}
\begin{tabular}{|c|c|c|c|c|c|c|c|}\hline\hline
$|S_{qc}, S_{\bar{q}\bar{c}}; S, L; J\rangle$                                &$M_Y(\rm{GeV})$     &Assignments       \\ \hline
$|0, 0; 0, 1; 1\rangle$\,(1P)                                                      &$4.24\pm0.09$       &$Y(4220/4260)$                  \\ \hline
$|0, 0; 0, 1; 1\rangle$\,(2P)
&$4.75\pm0.10$       &$Y(4750)$            \\ \hline

$|1, 1; 0, 1; 1\rangle$\,(1P)                                                      &$4.28\pm0.09$       &$Y(4220/4320)$             \\ \hline
$|1, 1; 0, 1; 1\rangle$\,(2P)                                                      &$4.81\pm0.10$       &               \\ \hline

$\frac{1}{\sqrt{2}}\left(|1, 0; 1, 1; 1\rangle+|0, 1; 1, 1; 1\rangle\right)$\,(1P) &$4.31\pm0.09$       &$Y(4320/4390)$               \\ \hline
$\frac{1}{\sqrt{2}}\left(|1, 0; 1, 1; 1\rangle+|0, 1; 1, 1; 1\rangle\right)$\,(2P) &$4.85\pm0.09$       &                \\ \hline

$|1, 1; 2, 1; 1\rangle$ \, (1P)                                                     &$4.33\pm0.09$       &$Y(4320/4390)$                 \\ \hline
$|1, 1; 2, 1; 1\rangle$ \, (2P)                                                     &$4.86\pm0.10$       &        \\ \hline
\end{tabular}
\end{center}
\caption{ The masses  of the vector tetraquark states and possible assignments, where the 1P and 2P denote the ground states and first radial excitations, respectively.   }\label{Assignment}
\end{table}

In Table \ref{Assignment}, we present the possible assignments of the vector tetraquark states based on the QCD sum rules. From the table, we can see explicitly that there is a room to accommodate the $Y(4750)$, i.e. the $Y(4220/4260)$ and $Y(4750)$ can be assigned as the ground state and first radial excited state of the $C\gamma_5\stackrel{\leftrightarrow}{\partial}_\mu \gamma_5C $ type tetraquark states with the $J^{PC}=1^{--}$, respectively.
We cannot identify  a particle unambiguously with the mass alone, we have to study the decays of those vector tetraquark candidates with the QCD sum rules to testify the assignments, it is our next work.

\section{Conclusion}
In the present work, we choose the diquark-antidiquark type four-quark currents with an explicit P-wave between the diquark and antidiquark pairs to investigate  the ground states and first radial excitations of the hidden-charm tetraquark states with the quantum numbers $J^{PC}=1^{--}$ via the QCD sum rules. Firstly, we take account of the ground states at the hadronic side only, and update the old  analysis by refitting the continuum threshold parameters and Borel parameters. Comparing  with the old calculations, we obtain better convergent behaviors  in the operator product expansion at the QCD side and uniform pole contributions $(40-60)\%$ at the hadronic side.   Secondly, we take account of both the ground states and first radial excitations, and focus on the first radial excitations, and obtain new predictions.  In calculations (also old calculations), we use  the modified energy scale formula $\mu=\sqrt{M^2_{X/Y/Z}-(4.1\,\rm{GeV})^2}$ to select  the suitable  energy scales of the QCD spectral densities so as to improve the convergent behavior of the operator product expansion and enhance the pole contributions.
  All in all, we explore and obtain the masses and pole residues of those 1P and 2P vector tetraquark states  in a systematic and consistent way. We obtain  the lowest vector tetraquark masses and make possible assignments of the existing $Y$ states, and observe that there indeed exists a hidden-charm  tetraquark   states with the $J^{PC}=1^{--}$ at the energy about $4.75\,\rm{GeV}$, which can account for the  BESIII data.

\section*{Acknowledgements}
This work is supported by the National Natural Science Foundation of
China with  Grant No.12175068.

\end{document}